\documentclass[a4paper,11pt]{article}
\usepackage{enumitem}

\usepackage[a4paper,
  left=2.5cm, right=2.5cm,
  top= 3cm, bottom=4cm]{geometry}
\usepackage{amsmath}
\usepackage{oldgerm}
\usepackage{amssymb}
\usepackage{bbm}
\usepackage[dvips]{graphicx}
\usepackage{epsfig}
\usepackage{color} 
\usepackage{cite}
\usepackage{epic}
\usepackage{hyperref}

\usepackage{empheq}

\numberwithin{equation}{section}

\usepackage[utf8]{inputenc}

\usepackage{graphicx}
\usepackage{mathrsfs}
\usepackage{amsmath}
\usepackage{amssymb}
\usepackage{braket} 
\usepackage{hyperref}
\usepackage{frontespizio}
\usepackage{comment}
\usepackage{epsfig}
\usepackage{float}
\usepackage{color}
\usepackage[title]{appendix}
\usepackage{multirow}
\usepackage{tikz}
\usetikzlibrary{decorations.pathmorphing}
\usepackage{varwidth}
\usepackage{tikz}
\usetikzlibrary{backgrounds,patterns,calc}
\usepackage{xparse}
\usepackage{caption}
\usepackage{enumitem}
\setlist[itemize]{leftmargin=*}

\usetikzlibrary{decorations.pathmorphing}

\tikzset{zigzag/.style={decorate, decoration=zigzag}}

\newcommand{\citep}{\cite}

\newcommand{\bea}{\begin{eqnarray}}
\newcommand{\eea}{\end{eqnarray}}
\newcommand{\be}{\begin{equation}}
\newcommand{\ee}{\end{equation}}
\newcommand{\ba}{\begin{align}}
\newcommand{\ea}{\end{align}}

\usepackage{subfigure}

\def\0{{\boldsymbol 0}}

\title{}
\author{}

\numberwithin{equation}{section}

\begin{document}
%
%
\begin{titlepage}

\flushright{\small{CERN-TH-2024-105}}

\vspace{-0.5cm} {\flushright {\small{.}}} \\

%
\begin{center}

\vspace{1.2truecm}

{\huge\bf{
Personal Reminiscences of Steven Weinberg\\[0.3cm]}}

\vspace{1.2truecm}

{\fontsize{10.5}{18}\selectfont
{\bf  Fernando Quevedo${}$
}}
\vspace{.5truecm}

{\small DAMTP,  Centre for Mathematical Sciences,  University of Cambridge,\\ Wilberforce Road,  Cambridge, CB3 0WA, UK}\\
{\small Theoretical Physics Department, CERN, CH-1211, Geneva 23, Switzerland
}
 
  \vskip 2.2cm

 \begin{abstract}
{\small{{My personal recollections are presented regarding my interactions with Steven Weinberg and the impact he had in my career from when I was his graduate student until the present.
 }}}
 \end{abstract}
\end{center}

\end{titlepage}







\renewcommand*{\thefootnote}{\arabic{footnote}}
\setcounter{footnote}{0}

\newpage


Steven Weinberg was already my hero when I was an undergraduate student in my home country Guatemala where I was fascinated by his First Three Minutes book \cite{Weinberg:1977ji}. It was hard for me to believe that physics was so powerful as to be able to describe, with some certainty, what happens in the very early universe.  Reading that this subject was treated by such a famous physicist (he had written the classic textbook on gravitation \cite{Weinberg:1972kfs} in the early 1970's and received the Nobel prize the same year I finished my undergraduate) made me more confident that the content of the book was based on serious physics rather than speculations.

I was extremely lucky to have started my PhD in 1981, just before Steve joined the University of Texas at Austin (UT). Coming from Guatemala, UT was the only place I could have hoped to be accepted in the US for my PhD, thanks to an initiative started by UT Professor Robert Little to support Central American physicists. I knew the high energy and relativity groups were already very strong at UT, and I was very pleased to have been accepted with the help of Robert Little and also Professor  Mel Oakes whom I have met in a school in Guatemala a few years before.

The first two years we have to take basic courses, called ‘the big seven' but we could sit in more advanced courses and Steve's Quantum Field Theory course was clearly the most popular. Attended by graduate students of all levels, postdocs and some faculty. It was a pleasure to see Bryce DeWitt sitting always on the first row and asking many difficult questions. I confess that most of the times I could not understand the details of their discussions. One instance that I remember vividly is when Steve was introducing the Fadeev-Popov formalism and Bryce did not seem to follow the argument. At some point Steve stopped and said: ‘but Bryce you invented this!'. The lectures were outstanding and Steve took them very seriously. Since this  was before his field theory books were written \cite{Weinberg:1995mt,Weinberg:1996kr,Weinberg:2000cr}, he distributed handwritten notes at the beginning of each lecture. 

One lecture clearly stood out. In early 1983 he came to the lecture room and before starting he smiled and said something like ‘I have just been informed from my contacts at CERN that the W-particle has been discovered and that there are hints also for the Z-particle'. With my fellow students we remember this moment as the  highlight of our time in UT. We do not agree on his exact words nor our immediate reaction, but we all agree the generalised feeling of excitement after his words. It is hard to describe how we all felt and we could only imagine how Steve was feeling that day. We were living science history in the making. For me, the only moment that matched this was to be  present at the CERN auditorium when the Higgs particle was announced, almost 30 years later, on the 4th of  July 2012. These moments are rare and touch deeply all of us that have dedicated our career to high energy physics.

After finishing the ‘big seven’ courses I managed to get the courage to ask Steve to be my PhD supervisor, he asked me if I knew Feynman diagrams and after my positive answer he asked me to read one  of  his more recent papers \cite{Weinberg:1982id} and discuss it with him after a week. An impossible task since I did not know supersymmetry at that time but I did my best to get an idea about that paper. Luckily for me, the week after  Steve was very busy  and suggested  me to talk with his postdoc, Joe Lykken (former deputy director of Fermilab) who was already working with him \cite{Hall:1983iz}. Joe assigned me a project which was an extension of his work with Steve and he helped me with the calculations, so I managed to pass my qualifying exam even though I was extremely nervous since the previous student of Steve had failed and left the field. We published our results \cite{Lykken:1983th} but before sending the preprint we needed  Steve's approval. Since he was traveling we sent him a copy and one day Steve's assistant (the legendary Adele Traverso) came to my office telling me ‘Professor Weinberg said that you messed it up and he wants to talk to you'. I checked the meaning of ‘messed-up' in the dictionary and then went to his office,  extremely nervous, to receive the phone call. He had found a typo in the paper (a low case $f$ instead of $F$ in one of the equations) and did not understand what this variable was. I explained it was only a typo and he was happy. I was born again.

After the qualifying exam I became a member of ‘Weinberg’s group’ and we met every week for the legendary ‘brown bag lunches' at his office. Each time one or two members of the group reported on some original work in progress or some review of recent papers. The atmosphere was tense for a PhD student but we learned a lot from Steve’s continuous interruptions to get to understand each subject in his own way and giving not only hard-time but also suggestions to the speaker for further research. This experience together with the interactions with all of Steve’s visitors and the new hires such as Willy Fischler and Joe Polchinski (and the other theorists who were already at UT, such as Cecil and Bryce DeWitt, Philip Candelas, Claudio Teitelboim (now Bunster), George Sudarshan, Duane Dicus, John Wheeler and the great group of postdocs and fellow students) made the PhD experience exceptionally rich.

During those years String Theory came to prominence and Steve and essentially the whole group moved to work on this direction. They were very exciting times and we felt UT was one of the top places in the world for the subject. This allowed me to finish a good PhD thesis under Steve’s supervision although much of my work I did with visitors like Paul Townsend and  fellow students such as Cliff Burgess and Anamar{\i}a Font \cite{Burgess:1985zz}. Joe Polchinski was one of my examiners and, being the first time he had to examine a thesis, he came with many questions to ask me after my standard presentation. Luckily for me Steve stopped Joe and told him that the thesis was good and we should finish fast the procedure.  He could reserve his questions and ask them later in private. I was very much relieved. 

Steve helped me then to move on my career. I am sure it was because of his special recommendation that I was given a nice postdoc position at CERN after UT and for the further jobs I got after, the fact that I was a student of Steve played a crucial role, until finishing with my professorship at Cambridge and even my ten years as director of ICTP. For this I will always be in debt to him.

During all the years since I left UT I was  in touch with him and we met in several occasions. Just to mention a few anecdotes:  

\begin{itemize}
\item In 1986 after finishing my PhD, I asked Steve if he would be willing to donate his collection of scientific journals to the Guatemalan national university (Universidad de San Carlos de Guatemala) which he kindly agreed. We needed only to get someone to pay for the transportation and Abdus Salam immediately agreed and arranged ICTP to cover the expenses. I used to call this as the last Weinberg-Salam collaboration.

\item 
For my 60th birthday conference Steve recorded a 5 minutes video to talk about me and to wish me a happy birthday. This video \cite{video} is now one of my greatest treasures. 

\item
On the exact day of the 50th anniversary of his Nobel prize paper (17 October 2017) I organised a special colloquium at ICTP that Steve kindly accepted to give with the title ‘Reminiscences of the Standard Model’ \cite{colloquium}

Since Steve had constantly rejected all attempts to have a conference in his honour\footnote{Whenever any of his former students or postdocs or the faculty members of his group proposed it, his reply was that it was too early to celebrate his achievements, giving the impression that the best was yet to come.},  I took this opportunity to give a more than 15 minutes introduction to the main scientific contributions of Steve. This of course does not make him justice but at least illustrates that what for normal top scientists an introduction of 1-5 minutes is enough to summarise their achievements, for Steve this was impossible and I needed more than 15 minutes for a ‘short' summary of his achievements. He was clearly the greatest physicist alive and took my introduction with good humour.

\item
A few years ago, after some email exchanges he told me he was talking with a used-car dealer and realised that his last name was Quevedo. Steve was curious if this last name was common in Spanish speaking countries and asked me about it. I don’t remember how I answer but always wondered what was Steve doing talking with a used-car dealer!

\item Even though I was Steve’s student we did not collaborate on scientific papers. He usually preferred to work by himself. However due to the sad departure of Joe Polchinski, we joined Raphael Bousso to write a scientific biography of Joe for the National Academy. I appreciated a very human aspect of Steve while writing this article \cite{Bousso:2020ukx}.

\item
The last time I met Steve in person was in February 2018 when I went to UT and gave a colloquium. He  introduced me very kindly  (he needed only 1-2 minutes of course). Here is a photograph that Professor Harry Swinney kindly took on that special occasion:

\begin{figure}[h!] 
\begin{center} 
\includegraphics[scale=0.5, trim = 1cm 2cm 0cm 5cm, clip]{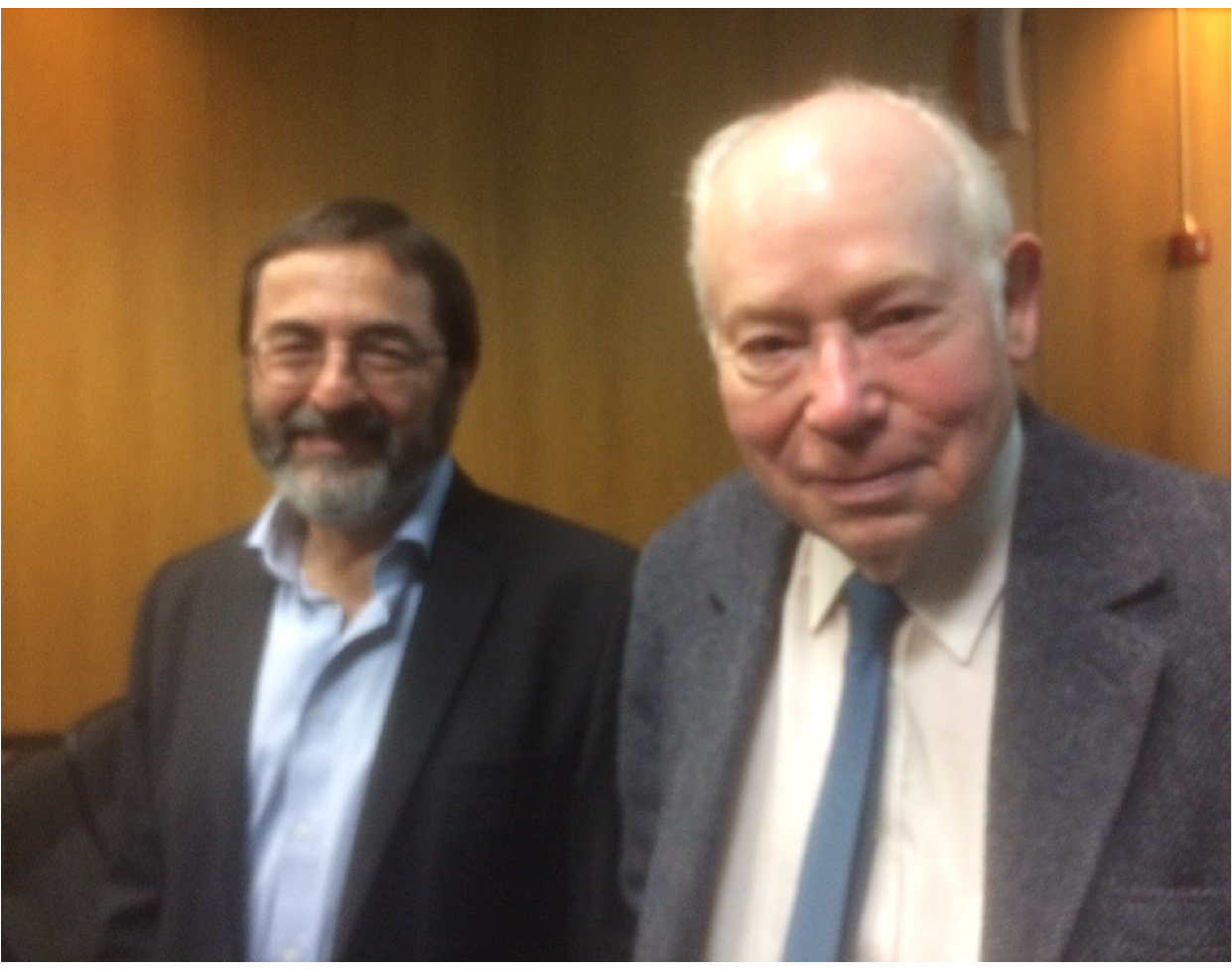}
\caption{\footnotesize Austin 2018.}
\end{center} 
\end{figure}

\end{itemize}

His many contributions to science, including his outstanding textbooks and popular science books, are well documented as well as his famous quotes on many aspects of science and related subjects. Here I would like to  share a couple of Steve’s quotes that may be less known:

During the question period after his colloquium at ICTP someone asked him how has he managed to get so much done in his career to what he replied: `very simple, I don’t go to church and I don’t ski so I have plenty of free time'. Finally, we were once having a conversation on religion with Don Page and he emphasised that you do not need to be religious to have high moral standards and said: ‘I know atheists that have the highest moral standards, for instance myself!'.

After  Steve passed away I was invited to give talks in different places summarising Steve's contributions to physics. I prepared a one hour talk concentrating essentially  on a small subclass of his renown papers (his almost 50 papers with more than 500 citations). I organised my presentation by 8 decades, since late 1950's until he passed away (I received the copy of his last book exactly the day he died). I did my best but it is actually impossible to summarise in a talk all his contributions, especially if his 8 textbooks and also 8 popular science books are included (see for instance \cite{APS_Texas})
\footnote{Mike Duff did an excellent job collecting the selected papers of Steve in \cite{Duff}.}.

Steve keeps influencing me as well as to most of the high energy physics community. As an example, during the past 4 years I lectured ‘The Standard Model' course at the Part III of the Cambridge Mathematical tripos\cite{SM}. After much thinking and preparation trying to give an original presentation to the subject I realised that at the end  I was essentially following Steve's  logical steps in his books (but not his notation). Starting with a chapter on the history of the Standard Model, something that he always insisted in his lectures. Followed by introducing particles as unitary representations of the Poincar\'e group and only after that introduce fields in order to describe local interactions among many particles that require assuming cluster decomposition on top of special relativity and quantum mechanics. Quoting him\cite{Weinberg:1995mt}  ‘If it turned out that some physical system could not be described by a quantum field theory, it would be  a sensation; if it turned out that the system did not obey the rules of quantum mechanics and relativity, it would be a cataclysm'.

In my lectures I  also use his soft theorems \cite{Weinberg:1965nx} to restrict to massless particles of helicity less than $2$ (and prove charge conservation and the equivalence principle using his theorems for massless particles of helicity $1$ and $2$  respectively).  I describe the power of effective field theories that he introduced over the years allowing us to consider non-renormalisable field theories, as gravity, in the same class as renormalisable theories except for their ultra violet completion. Using both the Fermi theory of weak interactions and chiral perturbation theory as examples. All this on top of introducing weak and strong interactions as gauge theories and gauge symmetries as redundancies in field theories to describe helicity $1$ particles consistent with Lorentz invariance and of course the Glashow-Weinberg-Salam model on which his less than 3 pages paper \cite{Weinberg:1967tq} may compete for the title of  the most confirmed predictive  information per written word. Predicting neutral currents, the $W^\pm$ bosons, the $Z$ boson, the Higgs particle.

For addressing ideas beyond the Standard Model I mention his gauge coupling unification work with Georgi and Quinn \cite{Georgi:1974yf}, his introduction of the axion field \cite{Weinberg:1977ma}, the famous Weinberg operator $H^2L^2/M$ to get neutrino masses and the whole SMEFT that relies on his EFT formalism, etc.\footnote{For recent discussions on his contributions to EFTs in his own words see \cite{Weinberg:2016kyd,Weinberg:2021exr}.} Essentially the whole course has Steve's legacy written on it.

It has been my greatest honour to have somehow my career and my life associated to such a giant scientific figure as Steve. It is impossible for me to exaggerate how much admiration I have for him.


\begin{thebibliography}{99}

\bibitem{Weinberg:1977ji}
S.~Weinberg,
``The First Three Minutes. A Modern View of the Origin of the Universe,''\\
    Basic Books (1977). Second edition (1993). 


\bibitem{Weinberg:1972kfs}
S.~Weinberg,
``Gravitation and Cosmology: Principles and Applications of the General Theory of Relativity,''
John Wiley and Sons, 1972,
ISBN 978-0-471-92567-5, 978-0-471-92567-5

\bibitem{Weinberg:1995mt}
S.~Weinberg,
``The Quantum theory of fields. Vol. 1: Foundations,''
Cambridge University Press, 2005,
ISBN 978-0-521-67053-1, 978-0-511-25204-4
doi:10.1017/CBO9781139644167

\bibitem{Weinberg:1996kr}
S.~Weinberg,
``The quantum theory of fields. Vol. 2: Modern applications,''
Cambridge University Press, 2013,
ISBN 978-1-139-63247-8, 978-0-521-67054-8, 978-0-521-55002-4
doi:10.1017/CBO9781139644174

\bibitem{Weinberg:2000cr}
S.~Weinberg,
``The quantum theory of fields. Vol. 3: Supersymmetry,''
Cambridge University Press, 2013,
ISBN 978-0-521-67055-5, 978-1-139-63263-8, 978-0-521-67055-5


\bibitem{Weinberg:1982id}
S.~Weinberg,
``Does Gravitation Resolve the Ambiguity Among Supersymmetry Vacua?,''
Phys. Rev. Lett. \textbf{48} (1982), 1776-1779
doi:10.1103/PhysRevLett.48.1776

\bibitem{Hall:1983iz}
L.~J.~Hall, J.~D.~Lykken and S.~Weinberg,
``Supergravity as the Messenger of Supersymmetry Breaking,''
Phys. Rev. D \textbf{27} (1983), 2359-2378
doi:10.1103/PhysRevD.27.2359

\bibitem{Lykken:1983th}
J.~D.~Lykken and F.~Quevedo,
``Stable Hierarchies in O'Raifertaigh Models Coupled to $N=1$ Supergravity,''
Phys. Rev. D \textbf{29} (1984), 293
doi:10.1103/PhysRevD.29.293

\bibitem{Burgess:1985zz}
C.~P.~Burgess, A.~Font and F.~Quevedo,
``Low-Energy Effective Action for the Superstring,''
Nucl. Phys. B \textbf{272} (1986), 661-676
doi:10.1016/0550-3213(86)90239-7

\bibitem{video}
https://mediacore.ictp.it/categories/00d285011e2b4ab25dc5f63a81ad9d9ec4a0291e-146303280/latest

\bibitem{colloquium}
https://indico.ictp.it/event/8413/

\bibitem{Bousso:2020ukx}
R.~Bousso, F.~Quevedo and S.~Weinberg,
``Joseph Polchinski: A Biographical Memoir,'' 
[arXiv:2002.02371 [physics.hist-ph]].
https://www.nasonline.org/publications/biographical-memoirs/memoir-pdfs/polchinski-josephb.pdf

\bibitem{APS_Texas}
https://meetings.aps.org/Meeting/TSF21/Session/L01.1

\bibitem{Duff}
M.J.~Duff, ``Steven Weinberg: Collected Papers" World Scientific (2024).

\bibitem{SM}
F.~Quevedo, A.~Scahchner 
``Cambridge Lectures on The Standard Model'' (to appear).

\bibitem{Weinberg:1965nx}
S.~Weinberg,
``Infrared photons and gravitons,''
Phys. Rev. \textbf{140} (1965), B516-B524
doi:10.1103/PhysRev.140.B516

\bibitem{Weinberg:1967tq}
S.~Weinberg,
``A Model of Leptons,''
Phys. Rev. Lett. \textbf{19} (1967), 1264-1266
doi:10.1103/PhysRevLett.19.1264

\bibitem{Georgi:1974yf}
H.~Georgi, H.~R.~Quinn and S.~Weinberg,
``Hierarchy of Interactions in Unified Gauge Theories,''
Phys. Rev. Lett. \textbf{33} (1974), 451-454
doi:10.1103/PhysRevLett.33.451

\bibitem{Weinberg:1977ma}
S.~Weinberg,
``A New Light Boson?,''
Phys. Rev. Lett. \textbf{40} (1978), 223-226
doi:10.1103/PhysRevLett.40.223

\bibitem{Weinberg:2021exr}
S.~Weinberg,
``On the Development of Effective Field Theory,''
Eur. Phys. J. H \textbf{46} (2021) no.1, 6
doi:10.1140/epjh/s13129-021-00004-x
[arXiv:2101.04241 [hep-th]].



\bibitem{Weinberg:2016kyd}
S.~Weinberg,
``Effective field theory, past and future,''
Int. J. Mod. Phys. A \textbf{31} (2016) no.06, 1630007
doi:10.1142/S0217751X16300076


\end{thebibliography}
\end{document}